\documentclass[useAMS,usenatbib,usegraphicx]{mn2e}

\title[Multi-periodic variations of BF Cyg]{Multi-periodic variations in the last 104 years light curve of the symbiotic star BF Cyg}
\author[E. Leibowitz and L.Formiggini]{E. Leibowitz$^{1}$\thanks{E-mail:
elia@wise.tau.ac.il} and L.Formiggini$^{1}$\thanks{{E-mail:
lili@wise.tau.ac.il}}\\
$^{1}$The Wise Observatory and the School of Physics and Astronomy 
\\ Raymond and Beverly Sackler Faculty of Exact Sciences
\\Tel Aviv University, Tel Aviv 69978, Israel}

\def\kms{$\rm km\, s^{-1}$}

\begin{document}

\date{Accepted 2005  . Received 2005  ; in original form 2005  }

\pagerange{\pageref{firstpage}--\pageref{lastpage}} \pubyear{2005}

\maketitle

\label{firstpage}

\begin{abstract}
We analyze a light curve of the symbiotic star BF Cyg, covering 114 years
of its photometric history. The star had a major outburst around the year
1894. Since then the mean optical brightness of the system is in steady
decline, reaching only in the last few years its pre-outburst value.
Superposed on this general decline are some 6 less intense
outbursts of 1-2 magnitude and duration of 2000-5000 days. We find a cycle
of ~6376 days, or possibly twice this period, in the occurrence of these
outbursts. We suggest that the origin of the system outbursts is in some
magnetic cycle in the outer layers of the giant star of the system, akin
to the less intense ~8000 days magnetic cycle of our Sun. We further find,
that in addition to its well known binary period of 757.3 days, BF Cyg
possesses also another photometric period of 798.8 days. This could be the
rotation period of the giant star of the system. If it is, the beat period
of these two periodicities, 14580 days, is the rotation period of a tidal
wave on the surface of the giant. A 4th period of 4436 days, the beat
period of the 14580 and the 6376 cycles is possibly also present in the
LC. We predict that BF Cyg will be at the peak of its next outburst around
the month of May in the year 2007. The newly discovered 798.8 days period explains 
the disappearance of the orbital modulation at some epochs in the light curve. 
The 757.3 oscillations will be damped again around the year 2013.   
\end{abstract}

\section{ Introduction}

Symbiotic stars (SS) are a class of variable stars consisting of a 
cool giant, a hotter object, either a hot subdwarf or a compact object, 
and an emission nebula.

The optical variability of symbiotics may take different forms and
time scales. One form is of cyclic variations, due to the varying aspects
of the revolving binary system, with or without an apparent eclipse in the
light curve (LC). Binary periods of SS are of the order of 1 to a few
years. Another type of variability has an explosive character, in the form
of a single outburst, as for symbiotic novae, or multiple events. The time
scales of these variations are quite long: the  decay times of the
outburst of symbiotic novae range between a few months to more than a
century. The cool giant in some symbiotic systems shows also intrinsic
variability such as radial pulsations of Mira-type.  

Variability of SS light on short time scale of minutes and hours has not
been reported much in the literature. Recently, however, such variations
on this time scale, reminiscing the flickering phenomenon in cataclysmic
variables, have been discovered in symbiotics (Sokoloski et al. 2001).

We have analysed anew the historical light curve of BF Cyg. This analysis 
resulted in the discovery of new elements in the long-term light curve of 
the star which may give us new clues on the nature of this system.  
 
In this paper  we present this analysis of the long-term  light
curve of BF Cyg, covering 114 years of observations. The characteristics  of
the symbiotic system BF Cyg are described in Section 2, and the data
sets used on our analysis are reviewed in Section 3.  In Section 4 we describe
the time series analysis and the periodicities detected.  In Section 5
we discuss the  physical interpretation of these periodicities.

\section {Brief description of BF Cyg}

BF Cyg is a very bright object (V $\simeq 9 $). This makes it a popular
target for observations on a wide wavelength range. This is also probably
the reason why the record of measurements of its magnitude goes back in
history for over 115 years (see below). Similar to many SS stars, and in
particular to the prototype Z And, the long-term light curve of BF Cyg
shows two kinds of optical variability. One is a regular periodic
modulation, with large changes in its amplitude. The other type of the
long range variability of this star is of explosive character, taking the
form of repeating outbursts (see Figure 1).

The cool component is a fairly normal M5 giant and the system is
classified as an S-type symbiotic with near-IR colors consistent with those 
of normal cool giants (Kenyon \& Fernandez-Castro, 1987; Munari et al. 1992). 
The IUE  ultraviolet continuum suggests the presence of a hot
subdwarf with a temperature $>60000 K$ (Gonz\'{a}les-Riestra, Cassatella \& Fernandez-Castro 1990).

Photometric variability with an apparent  modulation period of 754 days
was first noted by Jacchia (1941).
The system has been extensively studied in the optical and the ultraviolet 
wavelength ranges (Mikolajewska et al. 1989; Fernandez-Castro et al. 1990; 
Gonz\'{a}les-Riestra et al. 1990; Skopal et al. 1997). 
Its optical and
ultraviolet emission lines  vary in phase or/and in anti phase with the
photometric minima. The orbital nature  of this variation is confirmed by
infrared radial velocities data  and  the spectroscopic orbital period 
is 757.2, nearly identical to the photometric one (Fekel et al. 2001).

\section {The long-term light curve of BF Cyg}

We collected data from three large photometric measurement sets
retrievable for this system, in order to reconstruct its historical light
curve.

A regular photometric monitoring of BF Cyg for the years 1890-1940 is 
available from the Harvard plates (Jacchia, 1941). While few plates were 
collected in the first ten years, the data are more frequent after the
year 1900. 

A second data set is composed of  the photographic  measurements 
of Skopal et al. (1995).

A third large data set is the visual magnitude estimates collected by the
American Variable Stars Association (AAVSO). These are often daily
estimates of the magnitude of the system. For homogeneity we averaged the 
AAVSO data over a time interval of 24 days, similar to the sampling
interval  
of the Jacchia (1941) data set. The AAVSO at our hands is updated to Dec
2004.

\begin{figure*}
\includegraphics[width=130mm]{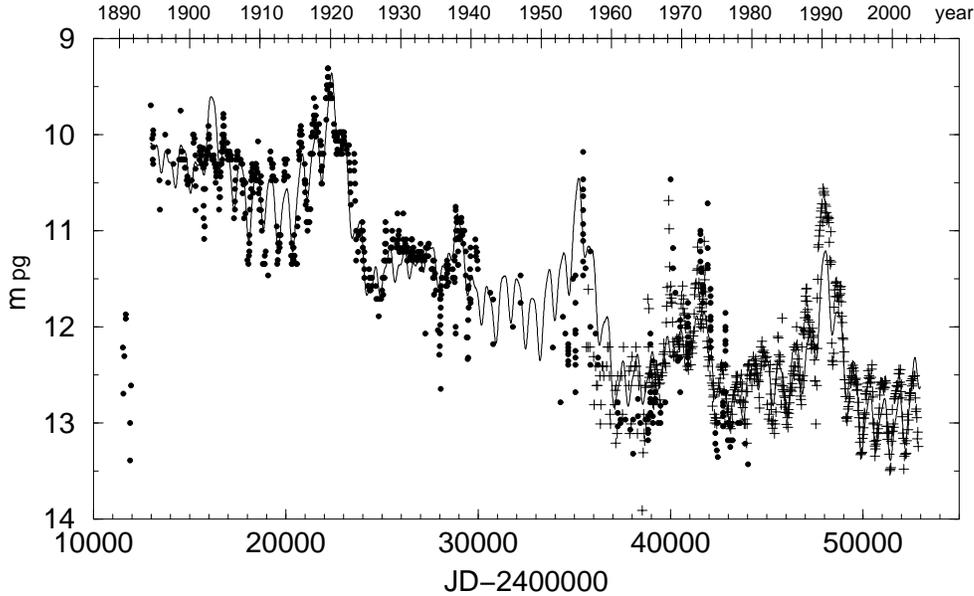}
\caption{A 114 years light curve of the symbiotic star BF Cyg, from the year   
 1890 up until Dec 2004.  Dots refer to m$_{pg}$ and crosses indicate  visual
data transformed to m$_{pg}$. The solid line is a 3d degree polynomial, and a 9
term harmonic wave based on 3 inependent periods, fitted to the data by least squares.
See text for further explanations.}
\end{figure*}

In Fig. 1 we present the long-term light curve (LC) of BF Cyg from 1890 up
to the present. In this figure, the AAVSO data have been scaled to the
photographic scale, by adding a factor .61 that takes into account the (B-V) 
color of the system at quiescence (Munari et al. 1992) and the transformation 
B=m$_{pg}$ +0.11 (Allen, 1973). This curve covers the photometric behavior 
of BF Cyg from 1890 to Dec 2004. There is however a considerable gap in the 
distribution of the data points between JD 2429986 and JD 2434486 which 
affects significantly the spectral window function of this time series 
(see Section 4.1) The solid line in the figure will be explained in 
Section 4.2. A sample of the data is shown in Table 1. Table 1 in
full  will be accessible only electronically.

\begin{tabular}{@{}llrcc@{}} \\
\multicolumn {2} {|c|}{\bf Table 1. The m $_{pg}$ or scaled m$_{v}$ data of Fig. 1}\\
\\
\\  
          Julian day   &  magnitude \\
\\                                                           
    2411547.9&  12.22 \\
    2411570.8&  12.70 \\ 
    2411616.4&  12.30 \\
    2411684.9&  11.87 \\
  \\
\end{tabular}  

The system underwent a dramatic 3.5 magnitudes brightening event around
the year 1894. This outburst is followed by slow fading of the star that
continues until the very present time. This behavior is reminiscent of
that of the small class of symbiotic novae, that have one single major
outburst recorded in their historical  light curve. Actually, there is
hardly a typical light curve 
of symbiotic novae, and the fading from the outburst shows a large variety
of behavior. It can be  gradual, such as in RR Tel or AG Peg, or
characterized by large light oscillations as in V1329 Cyg  or  by a deep
minimum as in PU Vul (Viotti, 1993).

In BF Cyg, a few major events of sudden brightening of the system by 1-2
magnitudes are superposed on the decline from
the major 1894 outburst. The duration of such an explosive event may last
a few years, and may include episodes of short term flares of
brightening by a  few tenths of magnitude lasting a few days. In
between outbursts, the system exhibits periods of relative quiescence that
last for a few years. In the last eight years BF Cyg is in one of these
quiescence states, while the system  is now back to the  
brightness level measured by Jacchia (1941) on the few Harvard patrol
plates obtained in the year 1890  before the large 1894 outburst.

A periodic oscillation of about 754 days was already recognized by
Jacchia (1941), with amplitude  $\sim $1 magnitude. Throughout the years
other values of the photometric orbital periodicity have been suggested by
various investigators. For example  P=757.3  days (Pucinskas, 1970), 
P=756.8  days (Mikolajewska et al. 1989), P=757.2  days (Fekel et al. 2001). 
As already noted  by Jacchia (1941), however, there were epochs in the history 
of the star, during which the binary modulation was very small, sometime 
nearly disappearing entirely from the LC.   

\section{Time Series Analysis}

The time series that we analyze in this work is the fading branch of the
light curve of BF Cyg that follows the outburst of the star in 1894. This
means that the few old measurements of the star magnitude prior to this
event have been discarded. We have detrended the series from the long-term
fading effect by subtracting from it a polynomial of 3d degree that was
fitted to the data by the least squares method. The dots (photographic 
magnitudes) and crosses (scaled visual magnitudes) in Figure 2 are
the detrended data points and they constitute the LC that we shall refer
to in this work. The solid line is a running mean of the data points over
the time interval of 757 days. The use of this line will be explained in
Section 4.2. 

\begin{figure}

\includegraphics[width=87mm]{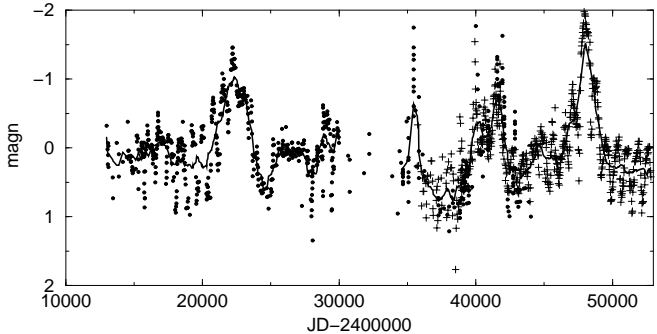}
\caption{Dots (mpg magnitudes) and crosses (visual transformed to mpg) represent 
the detrended LC of BF Cyg that is analysed in this work. Solid 
line is a running means curve over 757 days of the data points.}
\end{figure}

As a first step we computed the power spectrum (PS) of this LC (Scargle 1982). 
Figure 3 displays this PS in the frequency range corresponding to
the period interval between 40000 and 100 days. The upper bound is the
entire length of the LC. The lower bound is the period corresponding to
one half of the Nyquist frequency of the series. The insert shows the window 
function created by the non uniform sampling of the LC by the available measurements. 
Two groups of high peaks are clearly identified in the PS. One is at the red
end of the spectrum, around the frequency .00015 days$^{-1}$, and the other
around 0.0013 days$^{-1}$. We shall discuss them in turns in the next two
sections.      

\begin{figure}
\includegraphics[width=87mm]{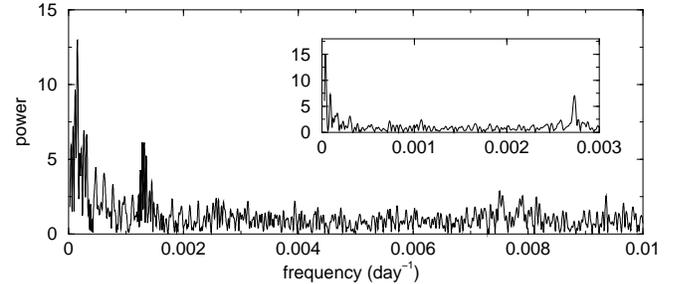}
\caption{Power spectrum of the light curve of BF Cyg shown in Figure 2. Insert is the window function.}
\end{figure}

\subsection{Periods between 1000 and 40000 days}

Figure 4 is a blowup of the PS shown in Figure 3 in the period range
between 1000 and 40000 days. We also considered a time series of 46 points
obtained from the observed LC by binning it into 46 bins of 757 days
width. A third LC that we also considered is the running mean shown in
Figure 2 as a solid line. These two LCs have identical PS to that shown in
Figure 4. Table 2 lists the frequencies, in days$^{-1}$, of the highest
peaks seen in Figure 4, as well as their corresponding periods, in days,
and the corresponding peak power. 

\begin{figure}
\includegraphics[width=3 inch]{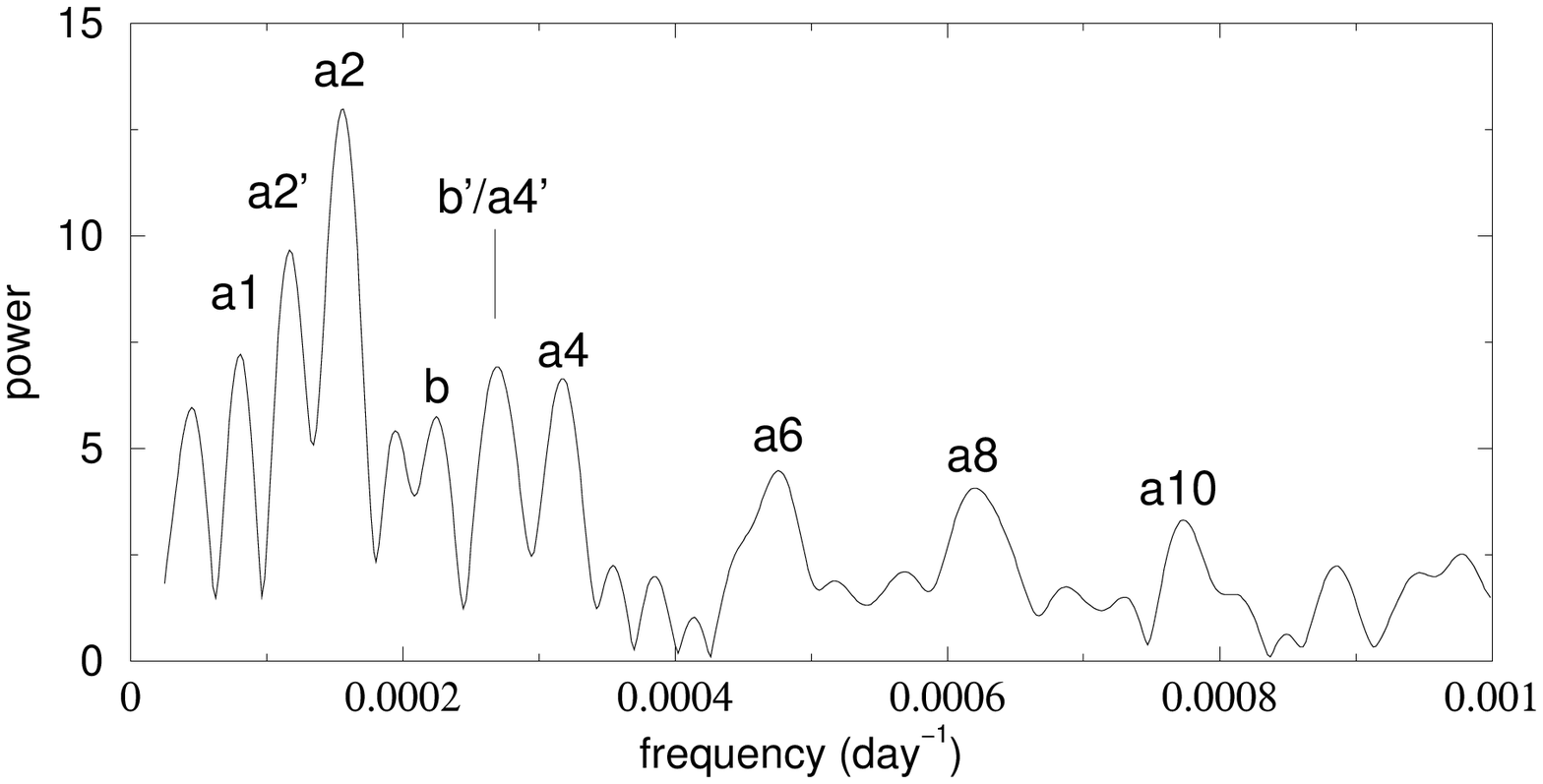}
\caption{Blowup of the red end of the power spectrum shown in Figure 3.}
\end{figure}

The highest peak (a2) in Figure 4, corresponding to the period of $\simeq$6400
days represents a periodicity in the outburst events of the star that is
also apparent in Figure 1. This periodicity is well established at a high level 
of statistical confidence as demonstrated in Appendix A (available electronically).

\begin{tabular}{@{}llrcc@{}} \\
\multicolumn {5} {|c|}{\bf Table 2. Peaks in the power spectrum }\\
\\
          Frequency   &        Period  &      Power       &  &\\
          days$^{-1}$  &         days  &  sigma unit & \\

\\                                                             
        4.4954 10$^{-5}$  &      22245   &     5.28   &     &  \\
         8.0871 10$^{-5}$ &      12365   &     6.39   &  a1 & P1   \\
          1.1679 10$^{-4}$&      8562.5  &     8.55   & a2$^{'}$ & alias\\
          1.5670 10$^{-4}$ &     6381.8  &    11.48   &   a2 & P1/2  \\
          1.9461 10$^{-4}$ &     5138.5  &     4.80   &     &         \\
          2.2454 10$^{-4}$ &     4453.6  &     5.08   &   b  & P4   \\
          2.7043 10$^{-4}$ &      3697.8 &     6.13   &  b'/a4$^{'}$&alias \\
          3.1832 10$^{-4}$ &     3141.5  &     5.88   &   a4  & P1/4\\
          4.7596 10$^{-4}$ &     2101.0  &     3.96   &   a6 & P1/6  \\
          6.1963 10$^{-4}$ &     1613.9  &     3.60   &   a8 & P1/8 \\
          7.7327 10$^{-4}$ &     1293.2  &     2.94   &   a10 & P1/10\\
         12.522  10$^{-4}$ &      798.61 &     3.79   &   p3 &P3 \\
         12.861  10$^{-4}$ &      777.55 &     5.44   &     &alias  \\
         13.200  10$^{-4}$ &      757.57 &     5.42   &   p2 &P2 \\
         13.539  10$^{-4}$ &      738.59 &      4.61  &     & alias \\
         14.517  10$^{-4}$ &      688.84 &     3.10   &      & \\
\\
\\
\end{tabular}

Nearly all other peaks in Figure 4 are related to this frequency in the
following way. The  peak denoted a1 corresponds to a period that is 
very nearly twice the period of peak
a2:  P1 $\sim$12400 days. Peaks a4, a6, a8 and a10 are the 4th, 6th, 8th and
10th  harmonics of the P1 periodicity. The peak marked a2' between a1 and
a2 is an alias of the dominant peak a2, due to the gap in the distribution
of the data points of the LC along the time axis seen in Figure 2 (Section 3). 
Its frequency is the sum of the frequency of a2 and of the frequency of the 
highest peak in the spectral window function, shown as insert in Figure 3 
(the high peak at the right hand side of the spectral window function corresponds
to the yearly cycle of 365 days). 
Indeed, the PS of each of the two subgroups of the data points that constitute 
the LC, when calculated separately, shows no trace of the a2' peak seen in Figure 4.

Peak b is quite prominent in the PS. It corresponds to the period P4$\sim$4450
days.  By itself it is not very significant statistically. We believe that
it is nonetheless a significant feature of the LC of this star. The reason
is that within the statistical uncertainties in the frequency values, its
frequency satisfies the equation 1/P4-2/P1=1/P5, where 2/P1 is the frequency of 
the highest peak a2, and P5$\sim$14700 days. The significance of the P5 periodicity 
will become apparent in Section 4.2. We show there that a similar relation 
1/P2-1/P3=1/P5 exists between P5 and two other prominent periodicities in the LC of the 
star, P2 and P3. The peak marked b'/a4' is an alias of its two neighbours 
on both sides, due to the gap in the data, as discussed above.  

Further evidence to the validity of our interpretation of the PS seen in Figure 3 is 
given in Appendix B (available electronically).

By least squares fitting, which will be described in the next section, we obtain the 
following best estimates of the values of the 3 periods P1, P4 and P5:  
P1=12750$\pm$400 days, P4= 4436$\pm$40 days, P5=14580$\pm$400 days.

\begin{figure}
\includegraphics[width=75mm]{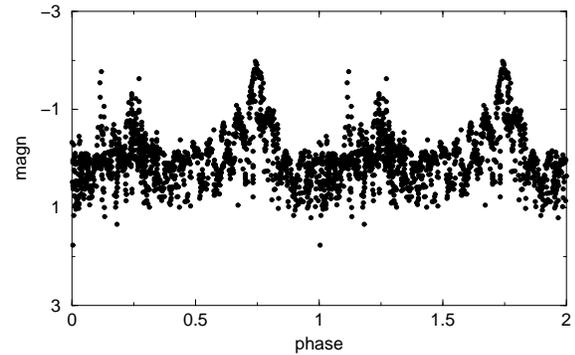}
\caption{Detrended light curve of BF Cyg folded onto the period  P1=12750 days.
 The cycle is exhibited twice.}
\end{figure}

Figure 5 presents the detrended LC of BF Cyg (Figure 2), folded onto the
period P1. The figure presents the cycle twice. The presence of the P1
peak in the PS is due to the apparent alternating height of the 6 recorded
successive outbursts, as presented by the two unequal maxima in the P1
cycle. 

The last recorded outburst reached its maximum light around JD
2448040. Based on our best fit values of the 3 periodicities P1, P2 and
P3 (see Section 4.2), we predict that BF Cyg will reach the formal maximum point 
of its next outburst around JD 2454236. In practice it means that the system will 
be found at the height of an outburst in mid-year ($\sim$May) 2007.  

\subsection{Periods between 200 and 1000 days}

Figure 6 is a blowup of the PS shown in Figure 3 in the period range 500 to
1000 days. Among the four peaks that dominate this figure, the one marked
p2 corresponds to the known binary period of the system P2=757 days. The peak 
marked p3 corresponds to the period P3=798 days. The two unmarked neighbors of 
the p2 peak correspond to the periods  777 and 738 days. They are aliases of 
the p2  peak, created by the gap in the distribution of the observed points, discussed
in Sections 3. Detailed explanation of this claim is given in appendix C (available electronically). 

\begin{figure}
\includegraphics[width=75mm]{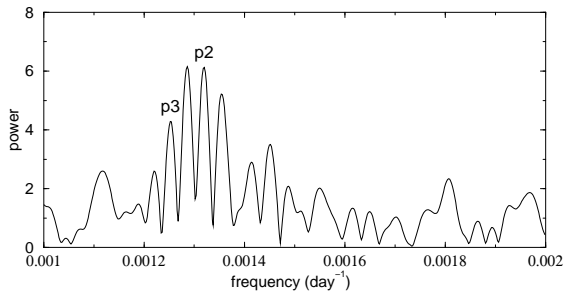}
\caption{Blowup of the section of the PS shown in Figure 3, containing the 
peaks around the binary orbital periodicity of the system.}
\end{figure}

Best estimates for the value of the periods P2 and P3, as well as of P1, P4 and P5, 
were obtained in the following way. We present the observed LC by a 9 term Fourier series, 
consisting of the P1 period with its first 5 higher even harmonics and the P4 periodicity, 
along with the P3 and the P2 periods. This presentation has only 3 free parameters, the values 
of P1, P2 and P3. The values of the other 6 periodicities are fixed by these 3. We find 
the value of these 3 parameters that give the 9 term series that fits best the observed LC in 
the least squares sense. The values of P1, P4 and P5 so obtained were given in Section 4.1. 
For P2 and P3 and their ephemeris we obtain:

Min(P2)=JD 2451443 $\pm$20 + (757.3 $\pm$0.9) $\times $E

Min(P3)=JD 2451531 $\pm$23 + (798.8 $\pm$1.4) $\times $E

Here the JD numbers are times of minimum light and the error figures
indicate half of a 95\% confidence intervals around the corresponding
values. They are computed by bootstrap (Efron and Tibshirani 1993) as detailed in 
Appendix D (available electronically).

Our P2 period is identical to the period already found  by Pucinskas (1970),
Min= JD 2415065 +757.3 $\times$E and confirmed  as the orbital one by
Fekel et al. (2001).

\subsubsection {Further discussion of periods P2 and P3 }

In order to  better analyze the photometric variability of the star in the
frequency range of the binary cycle 1/P2, we removed from the observed LC
all the variations on time scale shorter than the P2 period. This was done
by computing the running mean curve of the observed LC, averaging all
points within a window of width 757 days. This curve is shown as a solid
line in Figure 2. The dots in Figure 7 are the residuals of the observed
LC after removing the running mean curve from it. The solid line in Figure
7a represents a 2 term harmonic series of the P2 and P3 periods.

\begin{figure*}
\includegraphics[width=97mm]{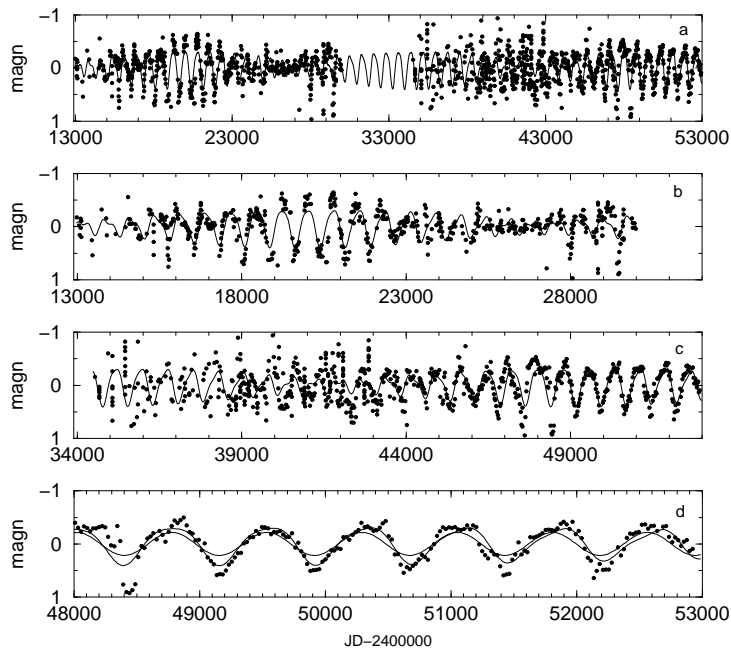}
\caption{Dots in all frames are the residuals of the observed light curve 
          after removing the running mean curve obtained with a 757 days wide 
          window, shown as solid line in Figure 2. (a) Solid line is the best  
          fitted 3 terms series with the periods P2, P3 and P2/2. (b) Zoom on 
          the first section of the plot in frame a. (c) Zoom on the second 
          section of plot a. (d) Blowup of the last 5000 days of the light 
          curve. The second thinner solid line is the best fitted harmonic wave
          with the Fekel's (2001) period of 757.2 days derived from radial 
          velocity data.}
\end{figure*}

In order to enable detailed examination, Figure 7b and 7c present,
respectively, zooms on the first and on the second subsections of the data
set of Figure 7a. Figure 7d is a blowup of the last 5000 days of the light
curve. The second, thin solid curve is the best fitted harmonic wave to the
data, with Fekel's et al. (2001) binary period derived from radial velocity
measurements.

Figure 7 demonstrates that the P2 orbital modulation, as well as the P3
oscillations, are present in the LC at quiescence states, as well as
during the outburst events. Except for the modulation due to the beat phenomenon, 
the structure and amplitude are independent of the luminosity of the system. 
In particular, the apparent binary variation has the same amplitude
at outburst maximum as during quiescence. It has also at the present epoch 
the same average magnitude as it had nearly a century ago, when the star 
was 3 magnitudes brighter.

The oscillations of the system with the P3 periodicity, simultaneously
with P2, explain in particular the epoch in the history of the star,
around JD 2427400, at which the 757 days oscillations have all but disappeared
from the light curve (see Figure 7b). This phenomenon was already noted by
Jacchia (1941)  more than 60 years ago. A similar episode of a nearly
disappearance of the orbital modulation is apparent again around JD
2442000. At this time the system was at the middle of one of its outburst 
events (Figure 2) and the intense activity of the star is masking the 
phenomenon to some extent, but it is still noticeable in Figure 7c. These 
two epochs of near disappearance of the 757 days variability are nodes of the 
beat between the P2 and P3 oscillations. The beat period is P5=14577, 
which is equal to the cycle of the beat between the 6376 and the 4436 periods
discussed in Section 4.1.

Based on our ephemeris of the two neighboring periods P2 and P3 we predict that 
the 757 variability of the system will be damped again around the year 2013.

The presence of the second period P3 in the LC in addition to P2 also provides  
an alternative explanation to the apparent shift in some of the minima of the 
binary cycle. Jacchia (1941) already noted that an O-C diagram indicates an 
apparent change in position of minima, which he thought to be correlated with 
the brightness of the star. Skopal (1998) reports on a similar effect in more 
recent parts of the LC. Skopal et al. (1997) report also on the presence of a 
possible secondary minimum in some of the 757 days cycles of the star. 

Figure 7  a,b and c show that most of these variations in the profile structure 
of the $\sim$757 days oscillations find a natural explanation in the ever changing
phase difference between the P2 and P3 frequencies. The figures show that not 
only the two phases  of damped oscillations are well explained by the contribution 
of the P3 periodicity. At epochs of quiescence of the star, when the $\sim$757 days
oscillations are not perturbed by an outburst activity, e.g. at 
JD 2415000 $<$ t $<$ JD 2418000 or JD 2448000 $<$ t $<$JD 2453000, the 2 periods 
LC fits the detailed structure of the observed oscillations quite well. It traces 
faithfully the minima of the oscillations and it also gives an apparent second 
minimum to some of the cycles. The interference of the two periods may
also explain the rather different values that investigators of this star
have derived from photometric time series at different epochs in the
history of the star (Jacchia 1941; Skopal 1998).  

In particular we point out, that due to this interference, the minimum
brightness times of individual cycles of the combined periodicity do not
fall necessarily at the minimum times of the binary period. Therefore, the
times of least brightness do not overlap exactly the times of inferior
conjunction of the giant star. One example is near the conjunction 
JD 2451395.2, determined by Fekel et al. (2001) from radial velocity
curves. From our synthetic LC we find that minimum light of the system was
reached only 48 days later, on JD 2451443. From the presently available
observed data, it is difficult to determine the time of minimum light with
the accuracy required to distinguish between these 2 dates.

Figure 7d 
shows the last 5000 days of the observed LC, along with our best fit 2
periods curve. The second, thin solid line is the best fit to the data of a
harmonic wave with Fekel's period of 757.2 days. Figure 7d seems to
indicate that the data support the notion that the later of the two dates,
resulting from our 2 periods presentation of the LC is closer to the true
minimum of the system.

 We may summarize this section with the claim that all the gross features
of the 1894-2004 LC of BF Cyg, following its outburst of 1894, are well
accounted for by a slow continuous decay, and 3 distinct periods P1=12750
days (+5 even harmonics), P2=757.3 days and P3=798.8 days. A
P4=14580 days period is also of significance although it is not presented
explicitly in the LC. It is the beat of the P2 and P3 periods and it is
manifested in the LC through the appearance of the P5=4436 d
periodicity. The solid line in Figure 1 is a plot of the 9 term series of
the 4 periods P1, P2, P3 and P4, with their corresponding harmonics,
superposed on a 3d degree polynomial representing the slow decay from the
1894 outburst.

\section{ Discussion}

\subsection{The outbursts cycle}

The 6 recorded outbursts of BF Cyg in the last 104 years occurred with a
constant time interval of $\sim$6376 days between them. Periodic or
quasi-periodic variations in the light of stars, with periods of thousands
of days and amplitudes of 2 or 3 magnitudes are known to exist in the class
of semi-regular giants (Mattei et al. 1988; Kiss et al. 1999).

 It seems, however, that the 6376 days periodic
variability of BF Cyg is of a different nature. The structure of the LC
shows the characteristics of outbursts, rather than of pulsations and so it 
is indeed interpreted by most researchers in the field 
(e.g. Gonz\'{a}les-Riestra et al. 1990; Skopal et al. 1997).
  
In an attempt to understand the origin of the outbursts phenomenon and the
nature of the clock that regulates their appearances, the clock of the
solar cycle, with its similar time constant of $\sim$8000 days, comes to
mind. We hypothesize that the origin of the cyclic outbursts of BF Cyg
lies in some magnetic activity, driven by a magnetic dynamo process in the
outer layers of the cool giant component of this stellar system. It is
known that the 11/22 year solar cycle modulates the mass flux of the solar
wind, among other measured parameters of the Sun. The basic process
originates in an interaction between differential rotation and convection
motions (Babcock  1961; Ulrich \& Boyden 2005). 

Solar-like cycle in Asymptotic Giant Branch stars
has been proposed by Soker (2000) as an explanation to the morphology of a
few planetary nebulae. Soker (2002) has also proposed that giant stars
that are members of symbiotic systems may harbor a magnetic dynamo
process.

The giant in BF Cyg may indeed possess one important ingredient of such a 
process, namely, a relatively fast rotation. It is classified as a M5 star
(Kenyon \& Fernandez-Castro 1987; M\H{u}rset \& Schmid 1999). 
No data are available for the rotation velocity of such late type stars in 
the De Medeiros et al. (2000) tables, but from the general trend of the velocity
distribution of cool stars, very small mean velocity (2-3 km/sec) is expected.  
Fekel et al (2001), on the other hand, derive a projected rotational velocity 
of the giant in BF Cyg of $\sim$4.5 \kms. 
This makes the giant of BF Cyg a fast rotator relative to field giants
of similar spectral type. We note also that differential rotation, another
important ingredient of the magnetic dynamo mechanism, has been recently
measured in some active K-type giants  (Weber, Strassmeier \& Washuettl 2005).

In the magnetic dynamo scenario, the repeating intensifications in the optical 
luminosity of the system that take the form of the outbursts, are due to
periodic enhancement of the stellar wind from the cool giant of the
system, regulated by this dynamo process. 
This results in an enhancement of the mass accretion rate onto the compact 
star of the system. The intense optical luminosity originates in the vicinity 
of the hot component, probably in a bloated gaseous shell around the WD star
(Munari 1989). 

We proposed in the past a similar solar-like cycle as an explanation for
the $\sim $ 8400 days ($\sim $23 years) cycle that we discovered in the LC
of another symbiotic star - Z And (Formiggini \& Leibowitz 1994). 
A comparison between BF Cyg and Z And reveals similarities between these
two systems and in their behavior both in quiescence and at outbursts. 
Their cool component  is a M5 star  (M\H{u}rset \& Schmid 1999) and 
the orbital period is almost the same: 758.8 days for Z And (Formiggini \& 
Leibowitz 1994) and 757.3 for BF Cyg (section 4.2).
Both systems belong to the classical symbiotics family, and during
outbursts their hot component seems to maintain a constant bolometric
luminosity while expanding in radius (Mikolajewska \& Kenyon 1992).

One difference between the 2 systems is that in addition to its sequence
of 6 "small" outbursts, BF Cyg underwent the dramatic 1894 event of large
outburst, quite distinct from the 6 that followed. No such event has been
recorded in the history of Z And. In view of the 1894 event, BF Cyg should
perhaps be classified as symbiotic nova, as already suggested by Skopal et al.
(1997). That event seems to be of the scale and nature that are different from
those of the small outbursts. The abrupt increase by 4 magnitudes in its
luminosity, and the slow decay lasting over 100 years afterwards, are
probably a signature of a thermonuclear runaway process, similar to the
events that characterize systems such as AG Peg, V1016 Cyg as symbiotic
novae (Mikolajeswka \& Kenyon 1992).

\subsection {The brightness oscillations}

Near the P2=757.3 days binary period of BF Cyg we discovered the period
P3=798.8 days. This could be the periodicity of oscillations of the M giant
star of this system. However, we consider this interpretation unlikely for
the following reason. The giant of BF Cyg is not a Mira type star. It
does not show modulations in the near infrared as for Mira. In the
(J-H) vs. (H-K) color diagram (Whitelock 1994) it does not lie in the
region occupied by Miras.

We suggest that the P3 periodicity is the rotation period of the giant of
the system. The rotation of the giant component of SS's was  already discussed
in the literature (e.g. Munari 1988). In general, the giant stars in symbiotic 
systems are assumed to rotate synchronously with their binary revolution. This assumption 
is based on theoretical considerations (Zahn 1977), taking into account the synchronization 
time scale expected from the values of the radii, masses and binary separations that are 
typical of symbiotics on one hand, and the estimated ages of these binary systems 
on the other. The period P3 is close enough to P2 and therefore would not be an unusual 
exception to this commonly believed rule.

Modulation of the star light at the rotation period of the giant is indeed expected in 
models explaining the binary photometric variations on the basis of the reflection 
effect (Kenyon 1986, Formiggini \& Leibowitz 1990). If the giant star is not strictly 
isotropic in its photometric characteristics, e.g. if there are dark spots on its
photosphere or if there are areas of different reflectivity on its
surface, rotation of this star will modulate the luminosity of the system
in the rotation frequency.

Modulations at the rotation frequency are expected also in models whereby the 
photometric binary variations are due to variation in the optical depth towards the 
main light source of the system that are coupled to the orbital revolution. In such models 
the WD and its close vicinity are seen by the observer through different regions of the
stellar wind of the giant star (Gonz\'{a}les-Riestra et al. 1990). The
stellar wind of the giant is not necessarily isotropic with respect to the
giant center. The non isotropic structure of the wind may also be coupled to the giant 
rotation through the effect of the stellar magnetic field. In such a case, the opacity 
toward the main light source of the system will vary at the giant rotation frequency, 
in addition to its variation at the binary frequency.

The P5=14580 days periodicity is the beat period of the binary orbital
period P2 and the P3 period. If the latter is  indeed the giant rotation period, P5 
is the period of the tidal wave that propagates in the outer layers of the giant, in the
coordinate system of the rotating star. We suggested above that a magnetic
dynamo process may be the driving mechanism responsible for the outbursts
cycle. It is not unreasonable to speculate that a tidal wave in
differentially rotating convecting layers may also modulate the magnetic
field of the star at the tidal wave frequency, in addition to the modulation by the
main cycle of the dynamo. The apparent P4 periodicity, if found to be real, may be 
the beat period of these two magnetic cycles in the outer layers of the giant.

Finally we note that as mentioned in Section 4.2.1 the structure and the amplitude of
the P2 and P3 oscillations are independent of the average luminosity of
the system. In particular it appears that the amplitudes of these modulations at maximum light
of outbursts are not significantly different from those during quiescence
states of the system.

This fact is consistent with the reflection interpretation of the binary modulation 
(Kenyon 1986, Formiggini \& Leibowitz 1990). To first approximation the reflection and 
the reprocessing of the hot component's radiation in the giant atmospheric layers, give 
rise to optical luminosity that is a given fraction of the hot component
luminosity. An increase in the later will result in the same relative
increase in the luminosity of the heated hemisphere of the giant, hence
the independence of the amplitude of the P2 variations, expressed in
magnitude units, on the luminosity of the system.

This independence is also consistent with the modulating optical depth
interpretation. A given optical depth of an absorbing layer reduces the
flux of the emerging radiation by a given percentage of the flux of the
radiation impinging on it, i.e. it has the same amplitude in magnitude
units.

\section{Summary}

We identify 3 independent periodicities in the LC of the last 104 years of
the symbiotic star BF Cyg. One is a 6376 days (or twice this
value) periodicity in the occurrence of outbursts on the decaying branch
of the luminosity of the system, following its major outburst in 1894. The
second is the well known binary period of the system of 757.3 days. The
third one is 798.8 days, which we suggest is the rotation period of the
giant star of this symbiotic system. A 4th period which is a beat of the
6370 days cycle with the beat period of the binary revolution and the giant
rotation period may also be present in the light curve.  

\section*{Acknowledgments}

       We acknowledge with thanks the variable star observations from the AAVSO 
International Database contributed by observers worldwide and used in this
research.

This research is supported by ISF - Israel Science Foundation of the
Israeli Academy of Sciences. 

\appendix
\section[]{Estimation of confidence level for P$\sim$6400 days}

We applied the bootstrap statistical test
(Efron \& Tibshirani, 1993 and see the Appendix D) on the 46 points LC
mentioned in Section 4.1, which averages all variations on time scale shorter than
757 days. We find that even in that LC of merely 46 data points, the
probability to obtain at random a distribution with a PS that has a peak
as high as the one seen in Figure 4 is less than 1/300. The periodicity
of $\simeq$6400 days in the occurrence of outbursts of BF Cyg is therefore  well
established at a high level of statistical confidence.

\section[]{Evidence for aliases around the frequency .00015 days$^{-1}$}
We also create an artificial LC by superposing the P1 period with its 5
higher even harmonics, and the P4 period on random noise with a  standard
deviation  that
is equal to that of the real data. We sample this LC at the times of the
real data. The PS of this time series is very similar to that of the real
data, including the high alias peaks a2' and b'/a4' seen in Figure 4.

\section[]{Evidence for aliases around the frequency .00013 days$^{-1}$}
We create an artificial LC
consisting of a 799 days and 757 days harmonic waves superposed on white noise,
and by sampling it at the times of the real LC. The PS of this time series
is similar to that of the real data. In particular, it contains the two
prominent unmarked peaks as in the real data, in addition to the high
peaks corresponding to the two planted periods p2 and p3.

We also computed separately the PS of each of the two subgroups of data
points that constitute the LC seen in Figure 2. The first group, LC1,
consists of all points at times smaller than JD 2429986 . The second group, LC2,
consists of all points with t$>$JD 2434486. Each of these 2 LCs covers less
than half the duration of the entire LC and has about half the number of
data points of the complete LC. The uncertainty in the frequency of the
highest peaks in the PS of LC1 and LC2 is therefore larger than in the
whole LC. The two highest peaks in the PS of LC1  correspond to the
periods 761 and 813 days, which within uncertainties of less than 2 sigma,
as determined by the bootstrap method (see Appendix D), are consistent with
the P2 and P3 periods. Likewise, the highest peaks in the PS of LC2
correspond to the periods 791 and 754 days, which are also consistent with
the P2 and P3 periodicities.

 The presence of the two periodicities in each
half of the data set is evidence to the reality of these two periods in
the LC of the star. Furthermore, the prominent unmarked peaks in the PS of
the full LC seen in Figure 6 are absent from the PS of either of the two
halves of the data set. This demonstrates that the unmarked high peaks are
indeed aliases of the p2 and p3 frequencies, created by the gap in the
data. 

\section[]{Estimation of the uncertainty  in the frequencies of peaks in a PS}

The estimate  of the uncertainty in the values of the components of the
vector F, consisting of frequencies of oscillations identified in an
observed light curve (t,y) is based on the statistical "bootstrap" technique 
(Efron \& Tibshirani 1993). 
Let W$_F$(t) be a harmonic series of the components of the frequency vector
F. We find by least squares the linear coefficients with which W$_F$ is best
fitted to the observed y values. For each point (t$_i$,y$_i$) we obtain the
residual d$_i$=y$_i$-W$_F$(t$_i$). We now create a pseudo-observed LC, by adding
to each synthetic value W$_F$(t$_i$) an element d$_j$ that is randomly selected
from the population d of the residuals. By least square fitting we find
the vector of frequencies F' for which the series W$_{F'}$(t) fits best the y'
values. After repeating this process N times, we have for each element F$_k$
of the frequency vector F, a sample of  N values of F'$_k$ . In the histogram of
all F'$_k$ values we find the narrowest interval  that contains the fraction
q of all N F'$_k$ values. We consider this as the full uncertainty interval
$\Delta$F$_k$ of statistical significance q around the frequency F$_k$. The
corresponding interval for the period is $\Delta$P$_k$=P$_k$ $^2$ $\times$ $\Delta$F$_k$.

\label{lastpage}

\end{document}